\documentstyle[12pt]{article}
\input{psfig}

\begin{document}
\flushbottom
\begin{titlepage}
\begin{flushright}
TU-451\\
July 1994
\end{flushright}

\vfill
 
\begin{center}{\Large\bf Looking up at the GUT/Planck World
}\end{center}

\vfill

\begin{center}{
Contribution to Proceedings of \\
the ICEPP Symposium \\
{\bf ``From LEP to the Planck World''} \\
University of Tokyo \\
17th-18th December 1992
}\end{center}

\vfill
 
\begin{center}{\large \bf Hitoshi~Murayama\footnote{Present Address: Theoretical Physics Group, Lawrence Berkeley Laboratory, University of California, Berkeley, CA 94720.}}\end{center}
\bigskip
\begin{center}{
Department of Physics\\
Tohoku University\\
Sendai, 980 Japan\\
(E-mail address murayama@tuhep.phys.tohoku.ac.jp)
}\end{center}
 
\vfill

\begin{abstract}
\noindent\normalsize
The importance of the mass spectroscopy of the superparticles is
emphasized. It will be shown that the gauge coupling constants give us
information on the GUT-scale mass spectrum once the superparticle masses
are known. The gaugino masses will provide us a model-independent test
of the grand unification irrespective of the symmetry breaking pattern.
The sfermion masses carry the information on the intermediate
symmetries, if present. Combining the mass spectrum, we will be able to
distinguish among the GUT models, and hopefully supergravity models also.
\end{abstract}

\vfill
\end{titlepage}

\section{\sc Statement}

I {\it assume}\/ that the supersymmetry will be discovered at future
colliders throughout this presentation. I do not argue that the
supersymmetry {\it must}\/ be discovered. I am not on a mission of the
supersymmetry; I am simply trying to convince you that there are lots of
interesting physics {\it after}\/ the discovery of the supersymmetry. Of
course, the discovery itself is very exciting. However, it seems even
more exciting to me that we will be able to ``peek'' the physics at the
GUT- or Planck-scale once the mass spectrum of the superparticles is
known.

The statement I wish to make in this talk is the following simple one.
\begin{quote}
The spectroscopy of the superparticle masses will provide us a unique
tool to verify/exclude/distinguish the GUT models.
\end{quote}

It is often stated to be an embarrassment of the supersymmetry that
there are more parameters than in the Minimal Standard Model.  It is
true that I cannot {\it predict}\/ the masses of the superparticles due
to the existence of many parameters. However, I argue that it is the
virtue of the supersymmetry that there are many more masses which are to
be measured experimentally. In the non-supersymmetric models, the only
clues to the GUT-scale physics is the gauge coupling constants and the
nucleon decay. On the contrary, there are many mass parameters which can
be measured at future colliders in supersymmetric models. These mass
parameters carry the information of the GUT or Planck world; whether
there is a unification or not, whether there is an intermediate scale,
and so on. For this purpose, precise measurements of the superparticle
masses are essential, and we have to design the future collider projects
and detectors so that such measurements will be possible.

\section{\sc What we learned from the Gauge Coupling Constants}

\begin{figure}
\centerline
{ \psfig{figure=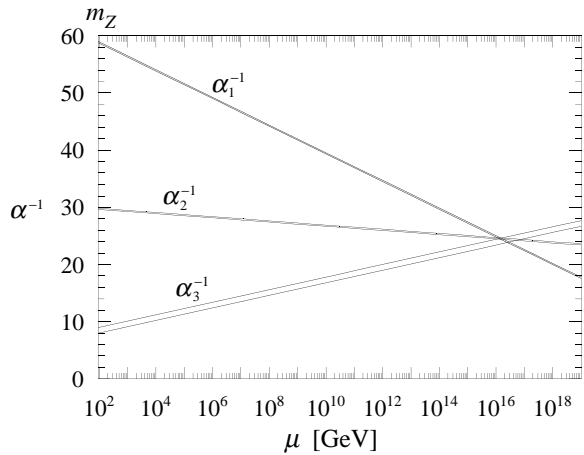,height=6cm}}
\caption[1]{\small The renormalization group flow of the gauge coupling
constants in the Minimal Supersymmetric Standard Model.}
\label{SU5}
\end{figure}

We have seen from the LEP data that the gauge coupling constants do meet
at a point if we assume the particle content of the minimal
supersymmetric standard model (MSSM). Though it can be still an
accident, we are encouraged to search for the supersymmetry at future
colliders.

It is important to summarize what we learned from the gauge coupling
constants measured at LEP. We often hear that the MSSM and $SU(5)$
SUSY-GUT is strongly supported by the LEP data. Indeed, the measured
gauge couplings do coincide around the scale $M_X \simeq 2 \times
10^{16}$~GeV with the MSSM particle content (Fig.~\ref{SU5}). Note,
however, that the gauge coupling constant unification alone does not
exclude the existence of light full $SU(5)$ multiplets. There are also
many reasons why the minimal $SU(5)$ model may not be the whole story.
It has a serious problem concerning the fine-tuning of the parameters to
keep Higgs doublets light while making their colored partners superheavy
(triplet-doublet splitting). The Yukawa coupling of the light species
are known not to obey the relations expected in the minimal model.
Furthermore, the cosmic baryon asymmetry which would be generated by the
minimal $SU(5)$ model is washed out due to the sphaleron effect since
the $B-L$ remains vanishing.

\begin{figure}
\centerline
{\psfig{figure=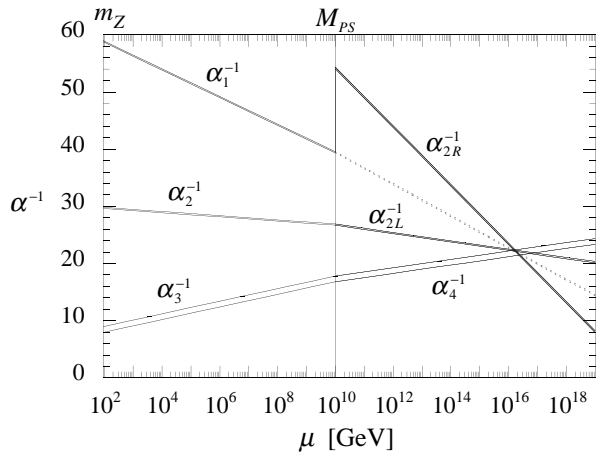,height=6cm}}
\caption[2]{\small The renormalization group flow of the gauge coupling
constants in the model with the chain symmetry breaking, $SO(10)
\rightarrow SU(4)_{PS} \times SU(2)_L \times SU(2)_R \rightarrow G_{SM}$
in Ref.~[2].}
\label{PS}
\end{figure}

It is interesting that there are $SO(10)$ models with an intermediate
scale consistent with the gauge coupling constants measured at LEP. The
first example was given by Deshpande {\it et al.}\/
\cite{Deshpande}, where the symmetry is broken as $SO(10)
\rightarrow SU(3)_C \times SU(2)_L \times SU(2)_R \times U(1)_{B-L}$,
and then down to the standard model gauge group $G_{SM}$. One can also
build a model with the Pati--Salam symmetry, $SO(10)
\rightarrow SU(4)_{PS} \times SU(2)_L \times SU(2)_R \rightarrow G_{SM}$
(Fig.~\ref{PS}) \cite{KMY}. In both cases, the intermediate scale cannot
be determined by the renormalization group of the gauge coupling
constants alone.

\begin{figure}
\centerline
{\psfig{file=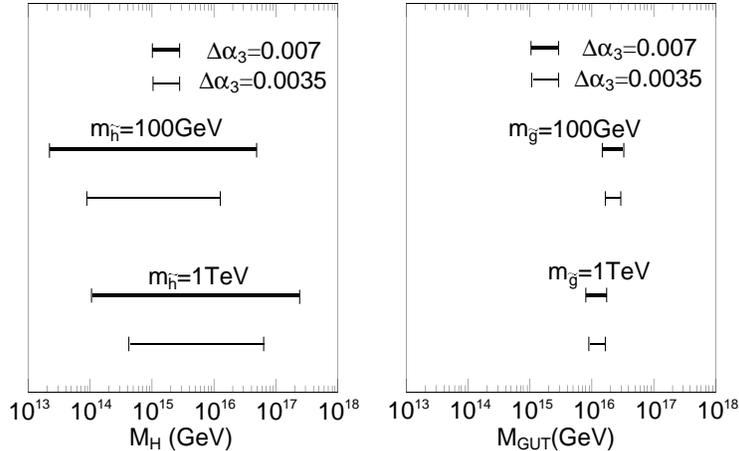,height=6cm}}
\caption[3]{\small The bounds on the GUT-scale mass spectrum from the gauge
coupling constants. The larger error bar corresponds to the present
accuracy $\Delta \alpha_s = 0.007$, while the smaller error bar to the
improved accuracy $\Delta \alpha_s = 0.0035$. The dependence on the
superparticle masses is also shown. The bound on $M_{H_C}$ depends
mainly on the higgsino mass $m_{\tilde{h}}$, while that on $M_{GUT}
\equiv (M_V^2 M_\Sigma)^{1/3}$ on the gluino mass $m_{\tilde{g}}$. }
\label{MHCfig}
\end{figure}

Nonetheless, it is interesting to see what we can learn from the gauge
coupling constants {\it assuming}\/ the minimal $SU(5)$ SUSY-GUT. The
particle content in the minimal $SU(5)$ model is quite simple, {\bf
24}-Higgs $\Sigma$ to break $SU(5)$ down to $G_{SM}$, {\bf 5} and {\bf
5}$^*$ Higgses to break electroweak symmetry down to $U(1)_{QED}$, and
the matter fields {\bf 5}$^*$ and {\bf 10} for each generations. Of
course, the gauge multiplet transforms as {\bf 24}. After the breaking
of the $SU(5)$ symmetry, the GUT-scale mass spectrum is characterized by
only three parameters, mass of the {\bf 24}-Higgses $M_\Sigma$, mass of
the superheavy gauge multiplet $M_V$, and mass of the colored Higgs
$M_{H_C}$. Using the renormalization group equations of the gauge
coupling constants, one can {\it measure}\/ these masses from the LEP
data \cite{HMY}. At the one-loop level of the renormalization group
equations, one obtains
\begin{eqnarray}
(3 \alpha_2^{-1} - 2 \alpha_3^{-1} - \alpha_1^{-1}) (m_Z)
	&=& \frac{1}{2\pi} \left\{ 
		\frac{12}{5} \, \ln \frac{M_H}{m_Z}
		- 2 \, \frac{m_{SUSY}}{m_Z} \right\},
			\label{MHC}
			\rule[-0.77cm]{0cm}{1.7cm}
			\\
(5 \alpha_1^{-1} - 3 \alpha_2^{-1} - 2 \alpha_3^{-1}) (m_Z)
	&=& \frac{1}{2\pi} \left\{
		12 \, \ln \frac{M_V^2 M_\Sigma}{m_Z^3}
		+ 8 \ln \frac{m_{SUSY}}{m_Z} \right\}.
		\label{MGUT}
\end{eqnarray}
Here, $m_{SUSY}$ stand for some weighted average of the superparticle
masses. Once the mass spectrum of the superparticle is measured, one can
determine $m_{SUSY}$ in the above formulae, and then extract the mass of
the colored Higgs $M_{H_C}$, and a combination of $M_V$ and $M_\Sigma$
from the renormalization group equations. The present bound is shown in
Fig.~\ref{MHCfig}, together with the expected limit if the precision of
$\alpha_s$ improves by a factor of two. In this way we can obtain an
upper bound on $M_{H_C}$ from the gauge coupling constants. On the other
hand, super Kamiokande experiment will put a lower bound on the
$M_{H_C}$ from the nucleon decay partial life time using the formula
(see Ref.~\cite{HMY2} for details and notations)
\begin{eqnarray}
\lefteqn{
\left.
\begin{array}{c@{=}c}
	\tau(p \rightarrow K^+ \bar{\nu}_\mu) & 0.62 \times 10^{31} \\
	\tau(n \rightarrow K^0 \bar{\nu}_\mu) & 0.35 \times 10^{31} \\
\end{array} \right\} 
}			& & \nonumber\\
& &	\times
	\left| 	\frac{\mbox{0.01~GeV}^3}{\beta} 
		\frac{M_H}{10^{16}\mbox{~GeV}}
		\frac{\sin 2\beta_H}{1+y^{tK}}
		\frac{(\mbox{10~TeV})^{-1}}{f(q,\,q) + f(q,\,l)} 
			\right|^2 \mbox{years}.
				\label{tauN}
\end{eqnarray}
Therefore, the precise determination of the gauge coupling constants
will be a crucial step to verify or rule out the minimal $SU(5)$
SUSY-GUT.

\section{\sc What we can learn from the Gaugino Masses} 

The gaugino masses satisfy a very simple renormalization group equation
at the one-loop level,\footnote{Recently, it was shown by Y.~Yamada
\cite{YY} that this relation is violated at the two-loop level. However,
its correction is of the order of a few percents, and irrelevant to the
discussion here.}
\begin{equation}
	\frac{d}{d\mu} \frac{M_i}{\alpha_i} =  0.
\end{equation}
Here, $M_i$ is the gaugino mass of the gauge group $i$, and $\alpha_i$
the corresponding gauge coupling constant. 

If the standard model gauge group is unified at some high energy scale,
then the boundary condition is that $M_i$ and $\alpha_i$ are common for
$SU(3)_C$, $SU(2)_L$ and $U(1)_Y$.\footnote{Note that the normalization
of the $U(1)$ gauge coupling constant may be model dependent. In all
known GUT models, however, the standard model is embedded into a group
larger than $SU(5)$. I will choose the $SU(5)$ normalization in this
talk.} Therefore, one expects in $SU(5)$ SUSY-GUT, 
\begin{equation}
\frac{M_1}{\alpha_1} =  
\frac{M_2}{\alpha_2} =  
\frac{M_3}{\alpha_3}  ,
\end{equation}
which is often referred to as the ``GUT-relation''. On the other hand,
the GUT-relation needs not to hold if the standard model gauge group is
not unified, as in some of the superstring models. The simplest example
is the flipped-$SU(5)$ model based on the gauge group $SU(5)_{\it
flipped} \times U(1)_{\it flipped}$ \cite{flipped}, where the $U(1)_{\it
flipped}$ gaugino may have different mass with the $SU(5)$ gaugino at
the unification scale,
\begin{equation}
\frac{M_1}{\alpha_1} \neq  
\frac{M_2}{\alpha_2} =  
\frac{M_3}{\alpha_3}  .
\end{equation}
The compactification of extra dimensions in the superstring theory may
directly lead to the standard model gauge group. Then all the three
$M_i/\alpha_i$ may differ with each other. Even if the gaugino masses
are ``universal'' at the Planck scale, the three gauge coupling
constants are different at the Planck scale phenomenologically (see
Fig.~\ref{SU5}), and hence do not satisfy the GUT-relation. The
deviation from the GUT-relation is at the 10 percent level, which can be
well distinguished at future $e^+ e^-$ colliders.  Therefore, the
gaugino mass will at least provide an experimental test of the $SU(5)$
SUSY-GUT.

One may suspect that the GUT-relation does not hold when there are
intermediate symmetry breakings between the GUT- and the weak-scales,
since the $U(1)_Y$ gaugino is a mixture of two or more gauginos which
have different masses and different coupling constants. However, one can
show that the GUT-relation of the gaugino mass {\it does}\/ hold even in
models with intermediate scales, as far as the standard model gauge
group is embedded into a simple group \cite{KMY}. For example, in the
model of Ref.~\cite{KMY}, the breaking pattern is $SO(10) \rightarrow
SU(4)_{PS} \times SU(2)_L \times SU(2)_R \rightarrow G_{SM}$, and
$U(1)_Y$ gaugino is a linear combination of the $SU(4)_{PS}$ and
$SU(2)_R$ gauginos. Though the mixing angles depend on the gauge
coupling constants of the groups and hence on the particle content above
the intermediate scale, the ratio $M_1/\alpha_1$ remains the same as
$M_2/\alpha_2$ and $M_3/\alpha_3$ irrespective of the gauge coupling
constants and the particle content.

One may also suspect that the threshold corrections at a symmetry
breaking scale may ruin the boundary condition that all the gaugino
masses should be the same. The threshold corrections may give large
effects when logarithms appear in the boundary condition of the
renormalization group equations. Indeed, there do appear logarithms in
the boundary condition of the gaugino masses depending on the mass
spectrum at the symmetry breaking scale. Fortunately, the logarithms
completely disappear when we take the ratio $M_i/\alpha_i$, since there
are exactly the same logarithms in the boundary condition of the gauge
coupling constants \cite{GHM}. The remaining threshold corrections are
only of the order of the ordinary radiative corrections, and hence of a
few percents.

In summary, the gaugino masses will tell us whether the standard model
gauge group is unified or not irrespective of the breaking pattern.

\section{\sc What we can learn from the Sfermion Masses} 

I will argue below that the sfermion masses will provide us the most
detailed information on the symmetry breaking pattern of the GUT-models
\cite{KMY}. 

The solution of the one-loop renormalization group equations of the
sfermion masses can be given explicitly as follows,
\begin{eqnarray}
m_a^2(\mu) &=& m_a^2 (\mu_0)
 + \sum_{i=1}^3 {2C_2(R_i^a) \over b_i}
(M_i(\mu_0)^2 - M_i(\mu)^2) ,
		\label{Sol}
\end{eqnarray}
where $i$ represents the gauge group, $a$ the species of the sfermion,
$C_2 (R_i^a)$ the second Casimir invariant of the gauge group
$i$ for the species $a$, and $b_i$'s are the coefficients of the
beta-functions defined by $\mu(\partial/\partial\mu) \alpha_i^{-1}
= - b_i/(2\pi)$.

Let us discuss the boundary conditions of the sfermion masses at a
symmetry breaking scale $M_{SB}$. Suppose the sfermion species $a, b, c,
\ldots$ belong to a single multiplet $A$ above $M_{SB}$. One naively
expects a
``unification'' of the sfermion masses at $M_{SB}$ as
\begin{equation}
m_a^2 (M_{SB}) = m_b^2 (M_{SB}) =m_c^2 (M_{SB}) = \cdots = m_A^2 (M_{SB}).
\end{equation}
However, unlike the gauge coupling constants or gaugino masses, there
are tree-level effects of so-called $D$-term contributions \cite{DHF}
which break the ``unification'' of the sfermion masses.  The boundary
conditions at the symmetry breaking scale $M_{SB}$ are,
\begin{equation}
m_a^2 (M_{SB}) = m_A^2 (M_{SB})
 + \sum_{I} {g_{I}}^2 Y_{I}^a Y_{I}^N \left(
 | \langle N \rangle |^2- | \langle \bar{N} \rangle |^2
		\right).  
		\label{smass}
\end{equation}
Here, the symmetry is assumed to be broken by the vacuum expectation
values $\langle N \rangle \simeq \langle \bar{N}
\rangle$,\footnote{Since a non-vanishing vacuum expectation values of
$D$-terms break supersymmetry, they are of the order of
$m_{SUSY}^2$, and the difference $\langle N\rangle - \langle
\bar{N}\rangle$ is of the order of $m_{SUSY}^2/M_{SB}$ when
$M_{SB} \gg m_{SUSY}$. } and $Y_I^a$ are charges of the broken generator
$I$ for the species $a$. The precise values of the $D$-terms are highly
model-dependent, and almost unpredictable. However, the coefficients
$Y_{I}^a Y_{I}^{N}$ are determined solely by the group theory. One can
also show that the $D$-terms do not depend on the choice of the Higgs
representation up to the overall normalization. Therefore, the relative
ratios between the $D$-term contributions to the various sfermions can
be predicted for each of the symmetry breaking patterns.

Now we focus on the nearest symmetry breaking above the weak-scale.
Assuming that the particle content is the MSSM one below the scale
$M_{SB}$ of the nearest gauge symmetry breaking above the weak scale, we
know the coefficients of the beta-functions $b_i$ as
\begin{equation}
	b_1 = \frac{33}{5}, \hspace{1cm}
	b_2 = 1, \hspace{1cm}
	b_3 = -3.
\end{equation}
The on-shell masses $m_{a,on}^2$ we will measure at future colliders are
different from the masses $m_a^2 (\mu)$ in the renormalization group
equations by the electroweak $D$-term,
\begin{eqnarray}
m_{a,on}^2 &=& m_a^2
 + M_Z^2(I_a^3-\sin^2\theta_W Q)\cos2\beta ,
		\label{mbar}
\end{eqnarray}
up to possible threshold effects at the weak scale, which are calculable
once the mass spectrum is known. If we observe gauginos, squarks,
sleptons and Higgs fields and measure their masses and $\tan\beta$, we
can obtain the values of sfermion masses $m_a^2(M_{SB})$ from
Eqs.~(\ref{Sol}) and ~(\ref{mbar}).  Therefore, we will be able to study
the sfermion mass spectrum at the nearest symmetry breaking scale
$M_{SB}$ from the TeV-scale experiments alone.

\begin{figure}
\centerline
{\psfig{file=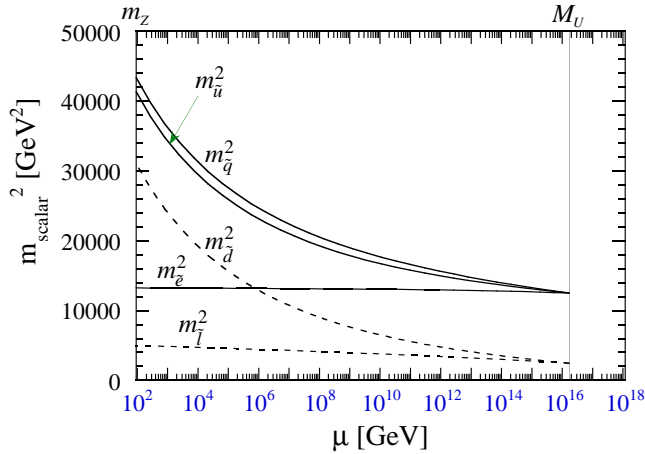,height=6cm}}
\caption[4]{\small Renormalization group flow of the sfermion masses in the
$SU(5)$ SUSY-GUT. The sfermion masses in the same multiplet should meet
at the same energy scale $M_U$ where the gauge coupling constants meet.}
\label{SU5scalar}
\end{figure}

Here, I will discuss only two specific examples. The first one is 
the $SU(5)$ model, whose nearest symmetry breaking scale is the
GUT-scale $M_U$ itself. The boundary conditions are simply
\begin{eqnarray}
&m_{\tilde{u}}^2 = m_{\tilde{q}}^2 = m_{\tilde{e}}^2 = m_{\bf 10}^2,&
	\label{SU5scalar1} \\
&m_{\tilde{d}}^2 = m_{\tilde{l}}^2 = m_{\bf 5^*}^2,& \label{SU5scalar2} 
\end{eqnarray}
at $M_U$. This model can be checked in the following manner.
First, we already know that the gauge coupling constants are consistent
with the expectations in the $SU(5)$ model. Second, once the
gaugino masses are measured, we can check whether they satisfy the
GUT-relation. At this stage, we still cannot distinguish between the 
$SU(5)$ model and other GUT-models. However, the sfermion masses provide
a highly non-trivial test of the $SU(5)$ model. The masses of
$\tilde{u}$, $\tilde{q}$ and $\tilde{e}$ should meet at a point; and the
energy scale should be the same as the scale where the gauge coupling
constants meet. The same should be true for $\tilde{l}$ and $\tilde{d}$.
A typical evolution of the sfermion masses is depicted in the
Fig.~\ref{SU5scalar}.

\begin{figure}
\centerline
{\psfig{file=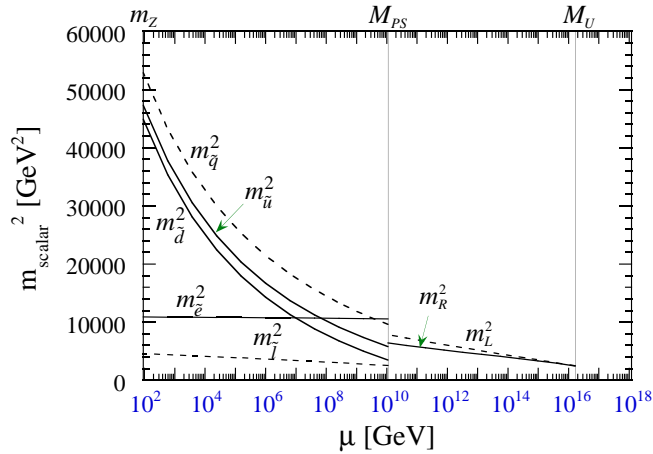,height=6cm}}
\caption[5]{\small Renormalization group flow of the sfermion masses in
the model with intermediate Pati-Salam symmetry. The sfermion masses
receive $D$-term contributions at $M_{PS}$. They have to satisfy the
relations (\ref{PS1}) and (\ref{PS2}) at $M_{PS}$. }
\label{PSscalar}
\end{figure}

The second example I discuss here is $SO(10)$ with intermediate
Pati-Salam symmetry \cite{KMY}. Even in this case, the gauge coupling
constants meet at the same scale as in the $SU(5)$ model (see
Fig.~\ref{PS}), and the gaugino masses satisfy the GUT-relation.  On the
other hand, sfermions do not have a simple mass spectrum as in the
$SU(5)$ model. Instead of Eqs.~(\ref{SU5scalar1}), (\ref{SU5scalar2}),
we have following two relations,
\begin{eqnarray}
m_{\tilde{q}}^2  - m_{\tilde{l}}^2 
	&=& m_{\tilde{e}}^2  - m_{\tilde{d}}^2 ,
		\label{PS1} \\
g_{2R}^2  (m_{\tilde{q}}^2 - m_{\tilde{l}}^2) 
	&=& g_4^2  (m_{\tilde{u}}^2 - m_{\tilde{d}}^2) ,
		\label{PS2}
\end{eqnarray}
at the scale $M_{PS}$ where the Pati-Salam symmetry is broken down to
the standard model gauge group. One of the relation should be used to
determine $M_{PS}$, and the other can be used to check the model. A
typical evolution of the sfermion masses is depicted in Fig.~\ref{PSscalar}.

\section{\sc Supergravity}

Once the mass spectrum of the superparticles is known, then probably all
the theorists will start seriously discussing the dynamics of the
supersymmetry breaking. Though it is usually believed that the
supersymmetry breaking occurs due to a dynamics around the Planck scale,
there is no compelling idea how the supersymmetry is broken at present. 
It may be interesting to recall that there are models of the
supersymmetry breaking which predict characteristic patterns of the
superparticle masses. The gaugino condensation in the hidden sector
generally leads to a baroque spectrum of sfermions, which may generate
unacceptably large flavor changing neutral currents.
Phenomenologically, models which predict ``universal'' scalar mass is
preferred. The no-scale model \cite{no-scale} predicts that all the
sfermion masses vanish at the Planck scale (and hence universal), and
the ``universal'' gaugino mass is the sole origin of the superparticle
masses. The dilaton-induced breaking \cite{dilaton} predicts that the
``universal'' gaugino mass is $\sqrt{3}$ times the ``universal'' scalar
mass.

Probably, the mass spectrum of the superparticles is the only clue to
the dynamics around the Planck scale. Since not all the theoretical
possibilities are classified at present, we cannot make definite
predictions yet. I am expecting the input from the experiments will
change this frustrated situation.

\section{\sc Prospects at JLC}

In regard of the previous discussions, one has to measure the masses of
the superparticles to compare with the predictions of various models. I
do not have a definite idea yet how accurately the masses should be
measured. For instance, the scalar masses should be measured quite
precisely if they are much heavier than the gaugino masses, since only
their {\it splittings}\/ carry the information of the symmetry breaking
pattern. If the sfermions are as light as the gauginos, then they
do not have to be measured so accurately. 

Anyway, the only future project which has potential to measure the
superparticle masses at the percent level is the $e^+ e^-$ linear
collider, such as JLC. It is a matter of luck whether we will be able to
find some of the superparticles at an $e^+ e^-$ linear collider of
$\sqrt{s} = 500$~GeV. But I strongly hope it will turn out to reality so
that we can distinguish among the GUT-models in the next 10 years.

Indeed, a detailed analysis has shown that the slepton and LSP masses
can be measured at a percent level; also the GUT-relation of $M_1$ and
$M_2$ can be checked at a similar accuracy \cite{JLC}. The mass
splitting between the right-handed and left-handed slepton depends
little on $\tan \beta$, and hence the difference between $m_{\bf 5^*}^2$
and $m_{\bf 10}^2$ can be measured very well.

Also, an $e^+ e^-$ linear collider with $\sqrt{s} = 500$~GeV can study
the top quark threshold region, which will enable us to determine the
QCD coupling constant $\alpha_s$ at a very high precision, $\Delta
\alpha_s \simeq 0.002$ \cite{Fujii}. As discussed in the section 2, the
precise value of $\alpha_s$ is useful to determine the GUT-scale mass
spectrum if we assume a specific GUT-model. 

One serious question is how we can measure precisely the mass of the
colored superparticles. As for the squarks, we will be able to produce
them in pairs if we can go up to sufficiently high energy with an $e^+
e^-$ machine. The problem is whether we can entangle the cascade decays.
For gluino, the situation is worse. We may be able to produce gluino
from the squark decay if kinematically allowed, which may allow us to
measure the gluino mass. If gluino is heavier than all of the squarks,
then we do not have any chance to produce gluino at an $e^+ e^-$
collider. The only strategy I can think of is to determine the
whole superparticle mass spectrum at an $e^+ e^-$ machine first, and then
compare the overall SSC/LHC data with the Monte Carlo. Anyway, it seems
to me challenging to measure the masses of colored superparticles.
Detailed and dedicated studies are needed.


\begin{thebibliography}{99} 

\bibitem{Deshpande} N.G.~Deshpande, E.~Keith, and T.G.~Rizzo, 
{\sl Phys. Rev. Lett.}\/ {\bf 70}, 3189 (1993).

\bibitem{KMY} Y.~Kawamura, H.~Murayama and M.~Yamaguchi, Tohoku
University preprint TU-439, ``Probing Symmetry-Breaking Pattern Using
Sfermion Masses'', June (1993), to appear in {\sl Phys. Lett.}\/ {\bf B}.

\bibitem{HMY} J.~Hisano, H.~Murayama, and T.~Yanagida, {\sl Phys. Rev.
Lett.}\/ {\bf 69}, 1014 (1992).

\bibitem{HMY2} J.~Hisano, H.~Murayama and T.~Yanagida, Tohoku University
preprint, ``Nucleon Decay in the Minimal Supersymmetric SU(5) Grand
Unification'', {\sl Nucl. Phys.}\/ {\bf B402}, 46 (1993).


\bibitem{YY} Y.~Yamada, {\sl Phys. Lett.}\/ {\bf B316}, 109 (1993).

\bibitem{flipped} I.~Antoniadis, J.~Ellis, J.S.~Hagelin, and
D.V.~Nanopoulos, {\sl Phys. Lett.}\/ {\bf 194B}, 231 (1987).

\bibitem{GHM} T.~Goto, J.~Hisano, and H.~Murayama,
Tohoku University preprint TU-423, ``Threshold Correction on Gaugino Masses at
Grand Unification Scale'', June (1993), to appear in {\sl Phys. Rev.}\/ {\bf D}.

\bibitem{DHF} M.~Drees, {\sl Phys. Lett.}\/ {\bf 181B}, 279 (1986);\\
J.S.~Hagelin and S.~Kelley, {\sl Nucl. Phys.}\/ {\bf
B342}, 95 (1990);\\
A.E.~Faraggi, J.S.~Hagelin, S.~Kelley, and D.V.~Nanopoulos, {\sl Phys.
Rev.}\/ {\bf D45} 3272 (1992).

\bibitem{no-scale} A.B.~Lahanas and D.V.~Nanopoulos, {\sl Phys. Rept.}\/
{\bf 145}, 1 (1987).

\bibitem{dilaton} J.~Louis and V.S.~Kaplunovsky, {\sl Phys. Lett.}\/
{\bf B306}, 269 (1993).

\bibitem{JLC} JLC-I, KEK Report 92-16, December (1992);\\
K.~Fujii, in this proceedings;\\ 
T.~Tsukamoto, K.~Fujii, H.~Murayama, M.~Yamaguchi, and Y.~Okada,
KEK-PREPRINT-93-146, 
``Precision Study of Supersymmetry at Future Linear $e^+ e^-$ Colliders'', 
November (1993), submitted to {\sl Phys. Rev.}\/ {\bf D}.

\bibitem{Fujii} K.~Fujii, in this proceedings;\\
Y.~Sumino, talk presented at Hawaii Linear Collider Workshop, April
(1993);\\
K.~Fujii, T.~Matsui, and Y.~Sumino, KEK-PREPRINT-93-125, October 1993,
submitted to {\sl Phys. Rev.}\/ {\bf D}.

\end{thebibliography}
\end{document}